\documentclass{article}
\usepackage{spconf,amsmath,graphicx,amsfonts,booktabs,fdsymbol,xcolor,multirow,xfrac}
\usepackage[inline]{enumitem}
\usepackage{hyperref}
\usepackage{cleveref}

\definecolor{xiaomi_gray}{HTML}{A9A9A9}

\title{Unified Keyword Spotting and Audio Tagging on Mobile Devices with Transformers}
\name{Heinrich Dinkel$^\dagger$, Yongqing Wang$^\dagger$, Zhiyong Yan$^\dagger$, Junbo Zhang and Yujun Wang\thanks{$^\dagger$ equal contribution.}}
\address{Xiaomi Corperation, Beijing, China}
\begin{document}
%
\maketitle
\begin{abstract}
Keyword spotting (KWS) is a core human-machine-interaction front-end task for most modern intelligent assistants.
Recently, a unified (UniKW-AT) framework has been proposed that adds additional capabilities in the form of audio tagging (AT) to a KWS model.
However, previous work did not consider the real-world deployment of a UniKW-AT model, where factors such as model size and inference speed are more important than performance alone.
This work introduces three mobile-device deployable models named Unified Transformers (UiT).
Our best model achieves an mAP of 34.09 on Audioset, and an accuracy of 97.76 on the public Google Speech Commands V1 dataset.
Further, we benchmark our proposed approaches on four mobile platforms, revealing that the proposed UiT models can achieve a speedup of 2 - 6 times against a competitive MobileNetV2.
\end{abstract}
\section{Introduction}
\label{sec:intro}

Keyword spotting (KWS) is currently a crucial front-end task for most intelligent voice assistants, which triggers the start of an interaction with the voice assistant if a user utters a specific keyphrase.
Further, Audio tagging (AT) is a task that aims to label specific audio content into sound event classes, e.g., the sound of a baby crying.
In the previous work~\cite{dinkel22_interspeech}, the authors have shown that modelling both tasks via a unified framework (UniKW-AT) is possible, significantly improving noise robustness without sacrificing KWS accuracy.

However, if UniKW-AT models were deployed in real-world scenarios, they would need to fulfil the same requirements as KWS models.
First, KWS models are situated on-device, and their size, i.e., the number of parameters, is limited.
Second, KWS models require a fast inference speed and a small floating point operations (FLOPs) footprint due to being ``always on''.
Third, the delay of a KWS model needs to be as low as possible, such that an assistant's wakeup is immediate.
While these requirements have already been researched thoroughly within the KWS community, to the best of our knowledge, no previous study has focused on lightweight and on-device computation of AT models.
This work aims to bridge the gap and introduce a multitude of transformer-based models for unified keyword spotting and audio tagging (UniKW-AT), which can satisfy the requirements mentioned above.

The benefit of such a UniKW-AT model is that outputs from the AT branch can be passed down further into the automatic speech recognition (ASR) pipeline, possibly enhancing robustness against noise. At the very least such a UniKW-AT model can act as a voice activity detector~\cite{dinkel2021voice,xu2021lightweight,novitasari22_interspeech}.

\subsection{Previous work}

Due to the practical importance of KWS, previous work is focused on decreasing a model's parameter size~\cite{choi19_interspeech}, increasing its inference speed~\cite{Mo2020} and reducing its false-acceptance rate~\cite{raju2018data}.
In terms of architecture, convolutional neural networks (CNNs) have been well researched within the community~\cite{sainath15b_interspeech,kim21l_interspeech,choi19_interspeech}, while more recently transformer-based models~\cite{berg21_interspeech,letr_icassp2022,wang2021wake,garg21_interspeech} and multi-layer perceptron (MLP) mixers~\cite{morshed2021attention,ng2022convmixer} have also been studied.
On the contrary, within the field of AT, most work focuses on improving the state-of-the-art performance on the well-known Audioset benchmark.
Works such as~\cite{Kong2020d} popularized CNNs, while~\cite{gong21b_interspeech} utilized transformers.
However, the majority of research within AT is solely focused on improving performance without consideration for the real-world deployment of these models.

\section{Unified keyword-spotting and audio tagging transformers}
\label{sec:approach}

This paper contributes a variety of transformer-based networks, further called \textit{u}nified \textit{t}ransformers (UiT), which aim to provide fast inference speed and reduce the model-parameter size and computational overhead while preserving KWS and AT performance.

\paragraph*{Unified Keyword Spotting and Audio Tagging}

\begin{figure}[t]
    \centering
    \includegraphics[width=0.9\linewidth]{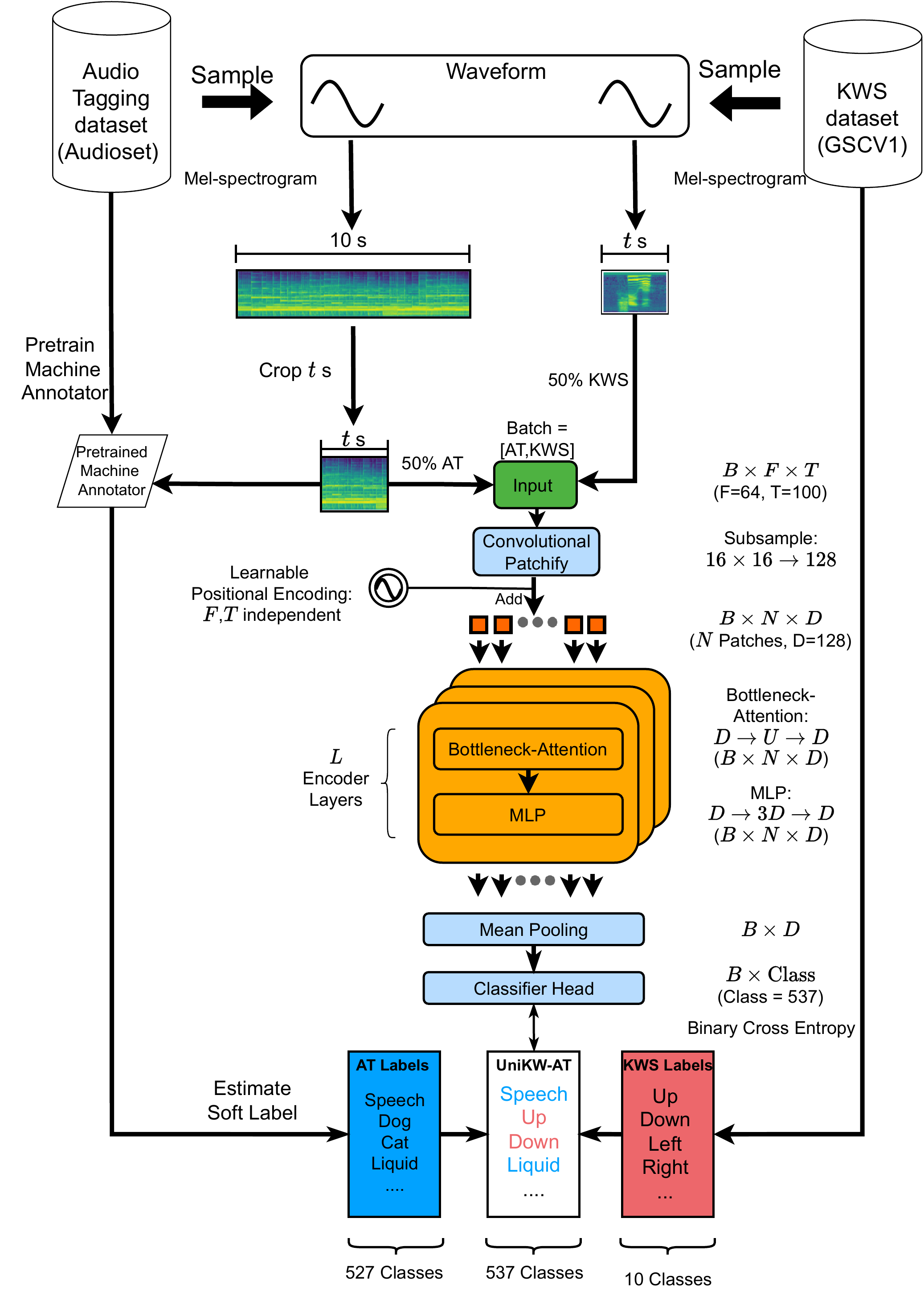}
    \caption{Depiction of the proposed UiT framework for training. We sample from an AT (Audioset) dataset and a KWS dataset (GSCV1). 
    Then we randomly crop the Audioset sample to match the KWS target sample length $t$. 
    Afterwards, a pre-trained model (MobileNetV2) is used to estimate these labels for these Audioset samples (pseudo-strong labels). During training, each batch is created using 50\% of AT and 50\% of KWS data. Training is done by optimizing the binary cross entropy (BCE) criterion.}
    \label{fig:framework}
\end{figure}

UniKW-AT has been proposed in~\cite{dinkel22_interspeech} and is modelled as follows.
Given the target KWS labelset $\mathbb{L}_{\text{KWS}}$ with $K$ keywords and an AT labelset $\mathbb{L}_{\text{AT}}$ with $C$ sound events, UniKW-AT merges both labelsets obtaining $\mathbb{L} = \mathbb{L}_{\text{KWS}} \cup \mathbb{L}_{\text{AT}}$.
Training samples from both KWS and AT datasets are randomly cropped to some target duration $t$ and the framework is optimized via the binary cross entropy (BCE) loss.
The entire training framework can be seen in \Cref{fig:framework}.

\paragraph*{Vision Transformers}

Transformers were first proposed for machine translation in~\cite{vaswani2017attention} and quickly became the state-of-the-art approach within the field of natural language processing (NLP).
Later in~\cite{Dosovitskiy_ViT}, the \textit{Vi}sion \textit{T}ransformer (ViT) has been proposed as an adaptation of transformers into the field of computer vision.
Then, ViT-based transformers were used in AT, where images were replaced with two-dimensional spectrograms~\cite{gong21b_interspeech,xu2022masked}.
The core idea of the ViT framework is the ``patchification'' operation, where an input image (here spectrogram) is first split into $N$ non-overlapping patches.
Each patch is of size $P=P_\mathtt{T} \times P_\mathtt{F}$ (time and frequency) and extracted via a convolution operation.
Then these patches are fed into a Transformer model consisting of $L$ identical blocks of a multi-head attention (MHA) layer followed by a multi-layer perceptron (MLP) layer.
A MHA layer computes:
\begin{align}
\label{eq:self_att}
    \mathbf{A} = \text{softmax}(\frac{\mathbf{QK}^T}{\sqrt{D}})\mathbf{V},
\end{align}
where $\mathbf{K} \in \mathbb{R}^{N\times D},\mathbf{Q} \in \mathbb{R}^{N\times D},\mathbf{V} \in \mathbb{R}^{N\times D}$ and are the key, query and value matrices obtained by a linear transformation $\mathbf{W}_j \in \mathbb{R}^{D\times D}, j \in \{K,Q,V\}$ of an input $\mathbf{X} \in \mathbb{R}^{N\times D}$.
The complexity of a single block is $\mathcal{O}(N^2 D + D^2 N)$, i.e., is quadratic for the model dimension $D$ and the number of patches $N$.

\subsection{Proposed Model}

We identify the embedding dimension $D$ and the number of patches $N$ as the primary reason for a large computational overhead and propose patch-reduction and bottleneck attention to reduce the complexity of our UiT model.

\paragraph*{Patch-reduction and Subsampling}

Common transformer models on Audioset are trained on a 10-s scale, utilize up to $N=N_{\mathtt{T}} \times N_{\mathtt{F}} = 1212$~\cite{xu2022masked,gong21b_interspeech} patches, leading to a high memory and computational burden for the transformer model.
Since time-frequency information is crucial for AT, we focus on reducing the amount of available time-patches $N_{\mathtt{T}}$ and fix the number of frequency patches $N_{\mathtt{F}}$ by feeding the model crops of length $t$ i.e., 1 s.
Limiting the context harms AT performance on Audioset since our model's training (1 s), and evaluation (10 s) duration is mismatched.
However pseudo strong labels (PSL)~\cite{dinkel2022_icassp} can partially alleviate the performance degradation by predicting fine-scale labels from a pretrained teacher (MobileNetV2) model.

The first layer within most CNN models, commonly known as stem, maps a $1\times T \times F$ spectrogram input to $C \times \sfrac{T}{2} \times \sfrac{F}{2}$, where $C \geq 16$~\cite{sandler2018mobilenetv2}, meaning that the overall memory required will be expanded by a factor of $\geq 4$.
Our approach uses a subsampling stem, directly reducing the memory requirement by mapping an input patch of size $P$ to a low-dimensional space $D$, where $D < P$.

\vspace{-3mm}
\paragraph*{Bottleneck attention}

Bottleneck attention (BN-A) focuses on reducing the dimension $D$ during the self-attention stage~\Cref{eq:self_att}.
Our intuition is that each respective patch-embedding within a spectrogram contains large amounts of redundant information.
Therefore, we propose using a BN-A approach, which reduces the dimension $D$ during self-attention to a lower space $U$, $U < D$.
We set $U=\frac{D}{4}$ for all our models.

\paragraph*{Architecture}

The proposed model architectures can be seen in \Cref{tab:archs}.
For all architectures we use $P = 256 = 16 \times 16$ patch-sizes, which sets the delay of our models to $16$ frames and use two heads for each self-attention operation.
This leads to $N_\mathtt{F}=4$ patches along the frequency axis and to $N_{\mathtt{T}} = 6$ patches for each input audio second.
We use ReLU as the default activation function due to its faster inference speed and lower computational complexity.
Similar to~\cite{letr_icassp2022}, we use a lower embedding dimension of $3D$ within the MLP, reducing the memory footprint.
\begin{table}[tbp]
    \centering
    \begin{tabular}{l||rrrrrr}
    \toprule
        Model &  $L$ & $D$ & MLP & \#Params & MFLops & $M_{pk}$\\
        \midrule
        UiT-XS & 12 & 128 & 384 & 1.5 M &  34 & 7.59 \\
        UiT-2XS & 6 & 128 & 384 & 0.8 M & 18 & 4.10 \\
        UiT-3XS & 4 & 128 & 384 &  574 k & 13 & 3.15 \\
        \bottomrule
    \end{tabular}
    \caption{Proposed UiT-based model architectures. The number of MFLOPs and the peak memory usage $M_{pk}$ (in MB) are calculated over 1 s.}
    \label{tab:archs}
\end{table}


\section{Experiments}
\label{sec:experiments}

\subsection{Datasets}
\label{ssec:dataset}

This work mainly uses the Google Speech Commands V1 (GSCV1)~\cite{warden2018speech} and Audioset~\cite{gemmeke2017audio} datasets.
We use the common 11 class subset of GSCV1 (V1-11), where the original 30 classes have been reduced to 10 common keywords: ``Yes'', ``No'', ``Up'', ``Down'', ``Left'', ``Right'', ``On'', ``Off'', ``Stop'', ``Go'' while the other 20 keywords are labeled as the AS label ``Speech'', where each sample is 1 s long.
We use the official training/validation/testing split containing 51,088/6,798/6,835 utterances, respectively.
As for AT training, we use the 60 h long balanced subset of AS containing 21,292 audio clips with a duration of at most 10 s per clip.

\paragraph*{Evaluation}

Evaluation is split between the KWS and AT subtasks.
AT evaluation uses the common evaluation subset of AS, containing 18,229 audio clips with an overall duration of 50 h.
KWS analysis primarily focuses on the GSCV1 dataset, which provides 2,567 target keywords and 4,268 non-target samples.
Note that for clips longer than the target length $t$, i.e., 10s, we split the input into chunks of length $t$ (i.e., 1 s), then feed these chunks into the model and average all output scores.

\subsection{Setup}
\label{ssec:setup}

Regarding front-end feature extraction, we use log-Mel spectrograms (LMS) with 64 bins extracted every 10 ms with a window of 32ms and a 16 kHz sampling rate. 
Our UiT transformer models use time- and frequency-independent learnable position embeddings.
We use random shifting, volume gain, and polarity inversion as augmentation methods in the waveform domain.
Having obtained a spectrogram, we augment the data using Specaugment~\cite{park19e_interspeech}.
Training runs with a batch size of 64 for at most 800 epochs using AdamW optimization~\cite{dettmers20218} with a linear warmup of 20 epochs to a starting learning rate of 0.001, which is then gradually decreased using cosine annealing.
We use mean average precision (mAP) during training as our primary validation metric.
Computing KWS accuracy for UniKW-AT requires post-processing since the model can predict multiple labels simultaneously, i.e., ``Keyword + Speech``.
Here we use a threshold of $\gamma = 0.2$, indicating the presence of a keyword.
The top-4 models achieving the highest mAP on the joint held-out validation dataset (GSCV1-valid and AS-balanced) are submitted for evaluation.
The neural network back-end is implemented in Pytorch~\cite{PaszkePytorch}.
To speed up the training procedure, we pretrain a single UiT-XS on the full Audioset using the masked Autoencoder approach~\cite{xu2022masked} and initialize all relevant layers for each model from this single checkpoint.
The source code is publicly available\footnote{\url{www.github.com/Richermans/UIT\_Mobile}}.

\section{Results}
\label{sec:results}

\subsection{Main results}
\label{ssec:main_results}

The core results of our work on the GSCV1 and Audioset can be seen in \Cref{tab:main_results}.
To illustrate the difficulty of training a UniKW-AT model, we also ran baseline experiments using a common TC-ResNet8~\cite{choi19_interspeech} model.
The results show that even though TC-ResNet8 can achieve excellent performance on GSCV1 (96.72 Acc), it fails to provide meaningful performance on AT (8.67 mAP).
Note that TC-ResNet8's performance on GSCV1 improves against the publicly reported result ($96.10 \rightarrow 96.72$) due to UniKW-AT training, where 60h of ``noise'' samples from Audioset enhance the model's robustness to false alarms.
As we can see, our proposed UiT-XS achieves competitive results compared to the previous MobileNetV2 (MBv2) based method for UniKW-AT, as well as other works in the literature regarding KWS and AT performance.
\begin{table}[tb]
    \centering
    \begin{tabular}{ll||rr}
    \toprule
    Approach & \#Params (M) & GSCV1  & AS \\
    \midrule
    TC-ResNet8~\cite{choi19_interspeech} &  0.06 & 96.10 & - \\
    NAS2~\cite{Mo2020} & 0.88 & 97.22 & - \\
    MEGA~\cite{ma2022mega} & 0.3 & 96.92 & - \\
    MatchBoxNet~\cite{majumdar20_interspeech} & 0.5 & 96.83 & - \\
    KWT-1~\cite{berg21_interspeech} & 0.6 & 97.05 & - \\
    LETR-128~\cite{letr_icassp2022} & 0.6 & 97.61 & - \\
    LETR-256~\cite{letr_icassp2022} & 1.1 & 97.85 & - \\
    KWT-2~\cite{berg21_interspeech} & 2.4 & 97.36 & -  \\
    KWT-3~\cite{berg21_interspeech} & 5.3 & 97.24 & -  \\
    Wav2KWS~\cite{seo2021wav2kws} & 225 & 97.90 & - \\
    \hline
    MBv2~\cite{gong2021psla} & 2.9 & - & 26.50 \\
    Eff-B0~\cite{gong2021psla} & 5.3 & - & 33.50 \\
    Eff-B2~\cite{gong2021psla} & 13.6 & - & 34.06 \\
    ResNet-50~\cite{gong2021psla} & 25.6 & - & 31.80 \\
    CNN14~\cite{Kong2020d} & 76 & - & 27.80 \\
    AudioMAE~\cite{xu2022masked} & 80 & - & {37.10} \\
    
    \hline
    \color{xiaomi_gray}  TC-ResNet8 & \color{xiaomi_gray} 0.1 &  \color{xiaomi_gray} 96.72 &  \color{xiaomi_gray} 8.67 \\
    MBv2~\cite{dinkel22_interspeech} & 2.9  & 97.53 &  33.42 \\
    MBv2$^\vardiamondsuit$~\cite{dinkel22_interspeech} & 2.9  & 97.53 &  32.51 \\
    \hline
    UiT-XS$^{\vardiamondsuit}$ & 1.5 &  97.76 & 34.09 \\
    UiT-2XS$^{\vardiamondsuit}$ & 0.8 & 97.31 & 32.21 \\ 
    UiT-3XS$^{\vardiamondsuit}$ & 0.6 & 97.18 & 30.97 \\ 
        \bottomrule
    \end{tabular}
    \caption{A comparison between our proposed UiT approaches against other works in literature. Results for GSCV1 use accuracy, while AS use mAP. Approaches denoted with $^{\vardiamondsuit}$ evaluate on 1 s chunks, influencing Audioset performance. Results with ``-'' means not available. }
    \label{tab:main_results}
\end{table}




\subsection{Inference latency on mobile hardware}
\label{ssec:inference_latency}

Here we measure the inference speed of our models using the PyTorch Mobile Benchmark Tool\footnote{\url{https://pytorch.org/tutorials/recipes/mobile_perf.html}}.
In \Cref{tab:inference_speed}, we display measured inference speed on four different mobile devices: 
We use two high-end Qualcomm Snapdragon chips, a Snapdragon 865 (SD865) and a Snapdragon 888 (SD888) and two mid-range chipsets, a MediaTek Helio G90T (G90T) and a Mediatek Dimensity 700 (MT700).
The results are compared to a TC-ResNet8 and an MBv2.
The MBv2 can be viewed as a baseline, representing a previous UniKW-AT approach, while TC-ResNet8 represents the speed requirement of a modern KWS system.

\begin{table}[htbp]
    \centering
    \begin{tabular}{l||rrrrr}
    \toprule
        Model &  SD865 &  SD888 &  G90T &  MT700 \\
        \midrule
        \color{xiaomi_gray} TC-ResNet8 &    \color{xiaomi_gray} 0.4 &    \color{xiaomi_gray} 0.4 &   \color{xiaomi_gray} 1.1 &   \color{xiaomi_gray} 1.1 \\
        MBv2      &    8.0 &    6.2 &  13.1 &   11.6 \\
        \hline
        Uit-XS    &    3.4 &    3.4 &   7.3 &    7.1 \\
        UiT-2XS   &    1.7 &    1.5 &   2.8 &    3.2 \\
        UiT-3XS   &    1.2 &    1.1 &   2.2 &    2.2 \\
        \bottomrule
    \end{tabular}
    \caption{Inference speed comparison on mobile system-on-a-chip (SoC) platforms, run on the central processing unit (CPU) with float32 precision and measured in ms. Each speed evaluation is accessed by first warming up the chip with ten warmup iterations followed by 1000 test trials for an input of length 1 s. }
    \label{tab:inference_speed}
\end{table}
As the latency results demonstrate, our proposed approach can achieve a speedup of over 2 times (8.0 $\rightarrow$ 3.4 ms) against an MBv2 when using UiT-XS while achieving a similar performance (see \Cref{tab:main_results}).
Even though UiT-2XS and UiT-3XS are slower than the baseline TC-ResNet8, they excel at AT (32.21/30.97 vs. 8.67  mAP, see \Cref{tab:main_results}).
Another important factor worth noting is that the baseline MBv2 has a delay of 320 ms, while our proposed models react within 160 ms.

\subsection{Ablation}
\label{ssec:ablations}

Here we present ablation studies focused on the proposed BN-A mechanism and the choice of a ReLU activation function in favour of a more common Gaussian Error Linear Unit (GeLU)~\cite{hendrycks2016gaussian}.
For simplicity, we only use the SD865 chipset for these tests.
The results can be seen in \Cref{tab:ablation}.
We observe that the proposed BN-A mechanism speeds up the inference time by at least 20\% ($4.1 \rightarrow 3.4$ ms) against standard self-attention without noticeable performance differences.
Moreover, while standard self-attention with GeLU can provide marginal performance boosts compared to the proposed BN-A + ReLU approach, they also significantly slow down the inference speed ($3.4 \rightarrow 5.7$ ms for UiT-XS) and increase the peak memory usage ($7.59 \rightarrow 10.28$ MB for UiT-XS), limiting their potential real-world use.

\begin{table}[htbp]
    \centering
    \small
    \begin{tabular}{ll||rr|rr}
    \toprule
         Model & Ablation & GSCV1 & AS & Speed & $M_{pk}$ \\
         \midrule
         \multirow{3}{*}{UiT-XS}  & Proposed & \textbf{97.76} & 34.09 &  \textbf{3.4} & \textbf{7.59}  \\
         & \, w/o BN-A &  97.75 & 33.76  &  4.1 & 10.28 \\
         & \, ReLU $\rightarrow$ GeLU &  97.69 &  \textbf{34.12}  &  5.7 & 10.28 \\
         \hline
         \multirow{3}{*}{UiT-2XS} & Proposed & \textbf{97.31}  & 32.21  &  \textbf{1.7} & \textbf{4.10} \\
         & \, w/o  BN-A &  96.94 & 32.30  &  2.2 & 5.44 \\
         & \, ReLU $\rightarrow$ GeLU  & 97.27 & \textbf{32.41}  & 2.9 & 5.44 \\
         \hline
         \multirow{3}{*}{UiT-3XS} & Proposed  & \textbf{97.18} & 30.97  &  \textbf{1.2} &  \textbf{3.15} \\
         & \, w/o BN-A  & 96.71 &  30.87  &  1.6 & 3.90 \\
        & \, ReLU $\rightarrow$ GeLU & 96.91  & \textbf{30.99}  & 2.1 & 3.90 \\
         \bottomrule
    \end{tabular}
    \caption{Ablation studies of the proposed model, where ``Speed'' represents measured ms and $M_{pk}$ is the peak memory requirement in MB. `w/o BN-A` represents the use of a standard self-attention mechanism, whereas `ReLU $\rightarrow$ GeLU` utilizes GeLU with standard self-attention. Results are evaluated on an input of 1 s. Best in bold.}
    \label{tab:ablation}
\end{table}

\section{Conclusion}
\label{sec:conclusion}

This paper proposes ViT-based transformer models (UiT) optimized for mobile device deployment of UniKW-AT.
Three lightweight UiT models are introduced, providing excellent performance for UniKW-AT coupled with fast inference speed, low peak memory usage and a small parameter footprint.
Our best model (UiT-XS) achieves an accuracy of 97.76 on the GSCV1 dataset and an mAP of 34.09 on Audioset, outperforming a competitive MBv2 baseline while having half the parameter footprint and a predominantly increased inference speed.


\vfill\pagebreak


\footnotesize
\bibliographystyle{IEEEbib}
\bibliography{refs}

\end{document}